# Adsorption of cytosine and aza derivatives of cytidine on Au single crystal surfaces

*Marianna Iakhnenko[1,2*], Vitaliy Feyer[3], Nataliya Tsud[4], Oksana Plekan[5], Feng Wang[7], Marawan Ahmed[7], Oleksandr Slobodyanyuk[2], Robert Acres[1], Vladimír Matolín[4] and Kevin C. Prince[1,6]*

[1] Sincrotrone Trieste S.C.p.A., in Area Science Park, Strada Statale 14, km 163.5, I-34012 Basovizza, Trieste, Italy.

[2] Taras Shevchenko National University of Kyiv, Faculty of Physics, Department of Experimental Physics, 64, Volodymyrs'ka St., 01601 Kyiv, Ukraine.

[3] Electronic Properties (PGI-6), Peter Grünberg Institute, Forschungszentrum Jülich GmbH, Leo-Brandt-Straße, 52428 Jülich, Germany,

[4] Charles University, Faculty of Mathematics and Physics, Department of Surface and Plasma Science, V Holešovičkách 2, 18000 Prague 8, Czech Republic.

[6] CNR-IOM Laboratorio TASC, Basovizza (Trieste), I-34149, Italy,

[7] eChemistry Laboratory, Faculty of Life and Social Sciences, Swinburne University of Technology, Hawthorn, Melbourne, Victoria 3122, Australia.

*Corresponding author. Tel.: +38 0506353696.
E-mail address: yakhnenko@gmail.com.

ABSTRACT. The adsorption of cytosine, 6-azacytosine, 6-azacytidine and 5-azacytidine on the Au(111) surface has been studied by soft X-ray photoelectron spectroscopy (XPS) and near edge X-ray absorption fine structure spectroscopy (NEXAFS). Monolayer films of these molecules were adsorbed on Au(111) from aqueous solution, and the nature of bonding with this surface has been determined. Cytosine was adsorbed from the gas phase by evaporation, as well as from solution, on both the Au(111) and Au(110) surfaces. The spectra are interpreted in the light of published calculations of the free molecules for cytosine and new ab initio calculations of the other molecules. Surface core level shifts of Au 4f imply that all of these compounds are chemisorbed. Cytosine adsorbs as a single tautomer, but in two chemical states with different surface-molecule bonding. For deposition in vacuum a flat lying molecular state bonded through the $N_{(3)}$ atom of the pyrimidine ring dominates, but a second state is also present. For deposition from solution, this second state dominates, and the molecular plane is no longer parallel to the surface. This state also bonds through the $N_{(3)}$ atom, but in addition interacts with the surface via the amino group. Two tautomers of 6-azacytosine were observed, and they and 6-azacytidine adsorb with similar geometries and chemical bonding via the azacytosine ring. The ribose ring does not appear to perturb the adsorption of azacytidine compared with azacytosine. The azacytosine ring is nearly but not perfectly parallel to the surface. 5-azacytidine adsorbs with the azacytosine ring very nearly parallel to the surface, as an imino tautomer.

## 1. Introduction

The interaction between biologically active molecules and metallic surfaces is an important topic both from applied and fundamental points of view and is relevant to many fields such as biophysics, medicine, biosensors, nanotechnology, etc. In the last few decades many studies have been focused on the binding of nucleobases and nucleosides to metal surfaces and metal colloids. Such investigations play an important role in the fundamental understanding of electron transfer through nucleic acids (DNA, RNA) and in future applications as a probe for nondestructive nucleic acid detection technologies in living organisms. It has become critical to understand how nucleic acids and their related biological building blocks interact with metal surfaces, and above all, with gold surfaces. Unlike many other metals, gold is stable in a biological environment, and gold nanoparticles have attracted huge interest for their applications as drug delivery systems, as sensors, and in other applications. These nanoparticles have some properties in common with bulk gold, and some that are uniquely determined by their small dimensions. In this work we take a first step towards understanding chemical bonding in such systems, by investigating well characterised bulk systems, in which we can determine the nature of chemical interactions between two dimensional gold surfaces and the molecules of interest. This work is intended as preparation for future studies of nanoparticles.

The anomalous pyrimidine base 6-azacytosine is a structural analogue of the canonical nucleobase cytosine in which the C-H group at the $C_{(6)}$ position of the ring is substituted by a nitrogen atom (Fig. 1). Similarly, the anomalous nucleoside 6-azacytidine (6-azaC) is a structural analogue of the canonical nucleoside cytidine, and consists of a pyrimidine derived ring, namely 6-azacytosine, and a ribose ring. In its isomer 5-azacytidine (5-azaC) the $C_{(5)}$-H group is exchanged with the $N_{(5)}$ atom in the pyrimidine ring. Both 6-azaC and 5-azaC nucleosides have a wide spectrum of biological properties that result in pharmacological activity, primarily antitumor and antiviral effects [1-7].

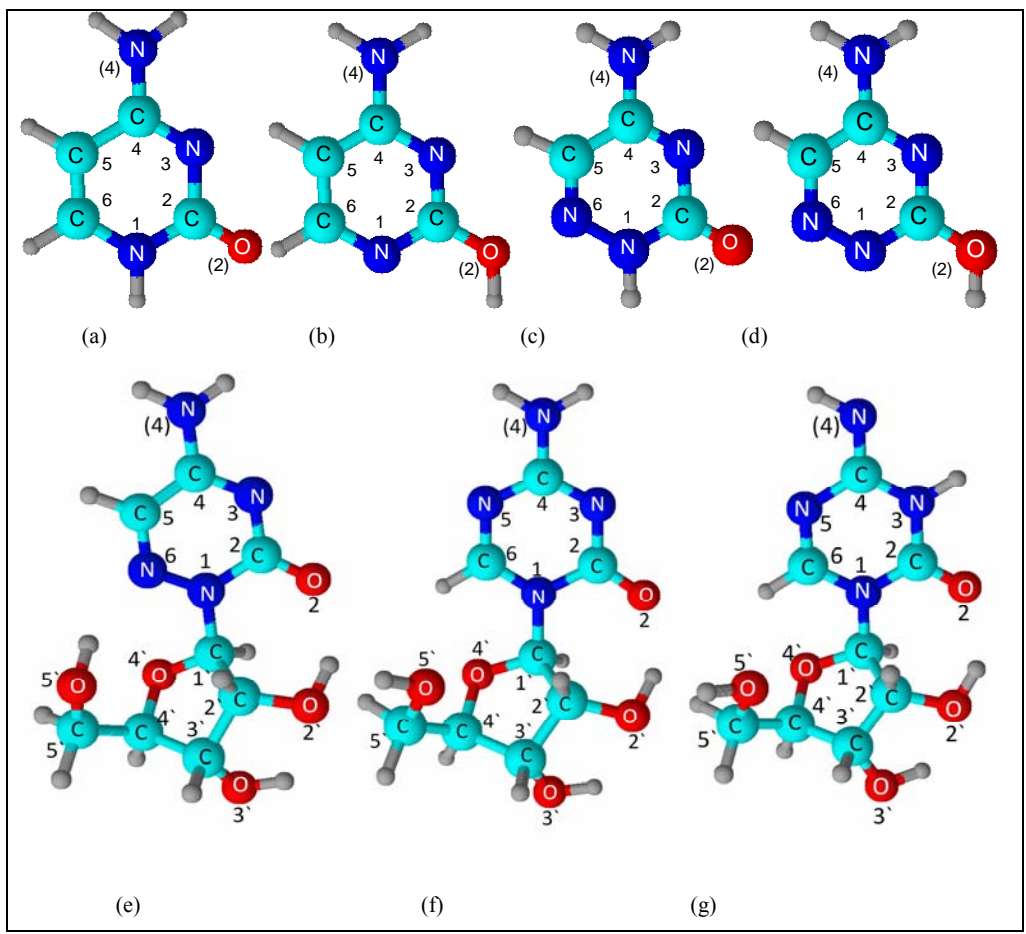

Figure 1: Structure of (a) cytosine, keto form, (b) cytosine, enol form, (c) 6-azacytosine, keto form, (d) 6-azacytosine, enol form, (e) 6-azacytidine, (f) 5-azacytidine and (g) 5-azacytidine, imino tautomer.

There have been extensive experimental [8-17] and theoretical [18-20] studies of biomolecules on gold and other Group 11 metal surfaces, to a large extent focused on amino acids, nucleobases, and their polymers - oligonucleotides, peptides, proteins, etc. We have chosen the present set of compounds based on (and including) the nucleobase cytosine as they are a representative set containing a standard nucleobase, as well as novel functional groups in the derivatives, such as one would expect to find in new pharmaceuticals, for example. The question we seek to answer is whether the modification of the molecular structure leads to a modification of the surface bonding. In addition, cytosine displays tautomerism in the gas phase [8] and in solution, and we wished to investigate whether this phenomenon is observable on surfaces. In this paper, we present an experimental and theoretical study of these compounds, beginning with cytosine and continuing with the derivative 6-azacytosine, and including two anomalous nucleosides. We compare calculations of the core level spectra of the isolated molecules with the experimental spectra of the adsorbed molecules. Calculations are available for cytosine [8], and we present new calculations for the other molecules.

The adsorption of cytosine on gold has been the subject of numerous studies using a wide range of techniques, such as STM, theoretical calculations, infrared and surface enhanced Raman spectroscopy and cyclic voltammetry [18-27]. Adsorption has been carried out in vacuum, in solution and in electrochemical cells, and on single crystals, thin films and nanoparticles of gold. For vacuum deposition, STM showed that cytosine was fluid at room temperature [19, 21], with rapid diffusion of the cytosine molecules (on the time scale of STM). On cooling, the adsorbates condensed into a disordered state, in which they were lying flat on the surface, and the bonding was dominated by intermolecular hydrogen bonds. A surface-molecule bonding energy of 0.1 eV (9.6 kJ/mol) was calculated, which is less than the calculated intermolecular hydrogen bond energy of ca. 15-40 kJ/mol between nucleic acid base pairs.

Various models of adsorption have been proposed, but there is no general consensus, in part because of the widely varying conditions used to deposit cytosine. The models include: flat lying molecules interacting weakly with the surface via dispersion forces, and with one another via hydrogen bonding [18,21]; chemisorption in the case of the closely related compound deoxycytidine on Au(111) [14]; chemisorbed molecules interacting via the $N_{(3)}$ atom, with the pyrimidine ring tilted at an angle of 50°, in one phase in an electrochemical cell [22,28]; tilted molecules interacting via the oxygen, $N_{(3)}$ and amino N atoms [29]; adsorption with a molecular orientation perpendicular to the surface, and along the [1$\bar{1}$0] direction [26]; on rough gold surfaces, adsorption either via $N_{(3)}$ or via the oxygen and $NH_2$ moiety [23-24].

There appears to be only one study employing core level photoemission [30], where the spectra were not published, and the XPS was used only to check stoichiometry. Demers et al [31,32] have measured the thermal desorption spectra of cytosine deposited from solution onto gold surfaces, and found molecular desorption with an enthalpy of 122 kJ/mol. From infrared data, they concluded that the molecule was not parallel to the surface, and desorption occurs from a single site of dispersed, non-aggregated cytosine. This experimental result is not consistent with the calculations of [18,21] in which the chief component of the adsorption energy is due to intermolecular hydrogen bonding.

Mansley et al [25] showed that adsorption of cytosine on Au(110) from solution prevents the reconstruction of the surface, and "freezes" it in a (1x1) configuration.

## 2. Methods

*2.1 Experimental methods and sample preparation.*

The experiments were performed at the Materials Science Beamline at the Elettra synchrotron light source in Trieste [33]. The beamline is equipped with a plane grating monochromator providing monochromatic synchrotron light in the energy range of 21-1000 eV, a Specs Phoibos 150 hemispherical electron energy analyzer, Low Energy Electron Diffraction (LEED) optics and a dual-anode X-ray source. The base pressure in the main chamber was in the $10^{-10}$ mbar range.

The Au 4f core level spectra were recorded at 120 eV of photon energy in normal emission geometry (incidence/emission angles of 60°/0°), and the total resolution (analyzer+beamline) was 0.15 eV. The C 1s and N 1s XPS spectra were collected in the same geometry, and the photon energy and total resolution were 500 and 0.45 eV respectively. The O 1s core level spectra were measured with the same multi-channel Phoibos analyzer using Mg $K\alpha$ radiation as the excitation source and the total energy resolution was 0.85 eV. Mg $K\alpha$ radiation was also used as the excitation source to measure C 1s, N 1s and O 1s spectra during preparation for the synchrotron radiation experiments. The binding energy (BE) was calibrated by measuring the Fermi edge. The NEXAFS spectra were taken at the N and O K edges using the nitrogen and oxygen KVV Auger yield, at normal (NI, 90º) and grazing (GI, 10º) incidence of the photon beam with respect to the surface. The energy resolution for the N and O K edge spectra was estimated to be 0.35 and 0.8 eV, respectively. The polarisation of light from the beamline has not been measured, but is believed to be between 80 and 90 % linear, as the source is a bending magnet. The raw NEXAFS data (N and O K edges) were normalized to the intensity of the photon beam, measured by means of a high transmission gold mesh and divided by corresponding spectra of the clean sample, recorded under identical conditions. The substrates were Au(111) and Au(110) discs of 10 mm diameter and 2 mm thickness supplied by MaTeck. They were cleaned *in situ* using standard procedures: cycles of $Ar^+$ sputtering (energy 1.0 keV), followed by annealing at 673-773 K. The surface order and cleanliness were monitored by LEED and XPS. Contaminants (such as C, N and O) were below the detection limits.

Cytosine was supplied by Sigma-Aldrich, and the samples of 6-azacytosine, 6-azacytidine and 5-azacytidine were synthesized in the Institute of Molecular Biology and Genetics of the National Academy of Science of Ukraine, Kyiv. These materials were used without any further purification.

Cytosine was deposited both by evaporation in vacuum at a temperature of 400 K and by deposition from solution. All other samples were prepared by adsorption from the liquid phase from saturated aqueous solutions, made from distilled water and the compound. The solution was prepared and all deposition steps were performed under a nitrogen atmosphere in a glove bag connected to a fast entry lock of the chamber. The Au(111) crystal, after preparation in UHV, was withdrawn from the chamber via the fast entry lock. A drop of the prepared solution was placed on the clean gold surface for 2 minutes. Then in the case of the cytosine and 6-azacytosine the surface was dried under a nitrogen gas flow (boil-off from liquid nitrogen), after which the gold crystal was transferred back into vacuum. The molecules adsorbed on the surface under these deposition conditions form saturated coverages (monolayers), and did not form multilayers. For the nucleoside samples (6-azacytidine, 5-azacytidine) the gold surface was additionally rinsed with distilled water and dried under nitrogen gas flow; this yielded layers several monolayers thick. A saturated coverage (defined as the monolayer) for these molecules was obtained using the following procedure. First we adsorbed multilayers on the surface and then flashed under vacuum from 325 K to 500 K in steps of 25 K; at 375-400 K the most weakly adsorbed species desorbed and on increasing the temperature, the intensity of C, N and O 1s did not change over a certain temperature interval. This coverage was defined as a monolayer and was further studied using synchrotron radiation. Under some circumstances, carbon contamination was observed, and was easily identifiable in the C 1s spectra.

Subsequent heating to 400 K of the saturated coverage of nucleoside samples did not change the areas of the C, N and O 1s XPS signals (measured with Mg $K\alpha$ radiation), while annealing to higher temperatures led to partial decomposition and desorption from the surface, which was indicated by a decrease of intensity, changes of stoichiometry and substantial changes in the shape of the XPS spectra.

We checked for radiation damage by monitoring the valence band and C 1s XPS spectra measured at 120 and 500 eV photon energy, respectively, and no spectral changes were observed after 1 hour of exposure to radiation. Thus the molecular films of these compounds are stable under our experimental conditions.

*2.2 Theoretical methods.*

The core level spectra of the free molecules were calculated for comparison with the surface spectra. We did not attempt to calculate the spectra of the adsorbed state, but used the theoretical gas phase spectra as reference data to understand whether relative shifts have occurred due to bonding with the surface. The theoretical methods used have been shown to give good agreement with experimental gas phase spectra [34,35] so differences from the theoretical spectra can be interpreted as due to surface effects. The methods have been described [34,35], and are summarized briefly here. The geometries of the free molecules were optimized using the B3LYP/6-311++G** model, followed by harmonic vibrational frequency calculations to ensure the optimized geometries are true minimum structures. The B3LYP/6-311++G** model is incorporated in the Gaussian G09 computational package [36]. Single point calculation was performed and based on the LB94/et-pVQZ model[37,38] to produce core vertical ionization energies. The single point calculations are produced using the Amsterdam Density Functional (ADF) computational chemistry package[39]

### 3. Results and discussion
*3.1. Theoretical structure results.*

Figure 2 shows the minimum energy conformations of the three compounds under study. The cytosine structures have been discussed previously[34,35], so that this section concentrates on 6-azacytosine (c and d), 6-azaC (e) and 5-azaC (f and g) and in the figure. The energy difference of the keto forms of 6-azaC (e) and 5-azaC (f) is small (113.63 kJ·mol$^{-1}$) although the dipole moment of 5-azaC given by 6.11 Debye whereas the 6-azaC is given by 6.97 Debye in our calculations. However, as shown in Figure 2, the intramolecular hydrogen bond network is quite different in the cytidine derivatives. For example, for both 5- and 6-azacytidine, the minimum energy conformations (e and f) includes a ribose OH…O$_{(2)}$ hydrogen bond, and an internal ribose hydrogen OH…O bond. For 6-azacytidine, there is also a ribose OH…N$_{(6)}$ hydrogen bond, which cannot be formed in 5-azacytidine for steric reasons. The ribose rings are not coplanar with the azacytosine rings. The calculated structure of 6-azacytidine is in agreement with the measured x-ray crystallographic structure, thus verifying the good quality of the calculations; we are not aware of structural data for 5-azacytidine. Full details of the structural parameters and nomenclature are given in the Supplementary Information.

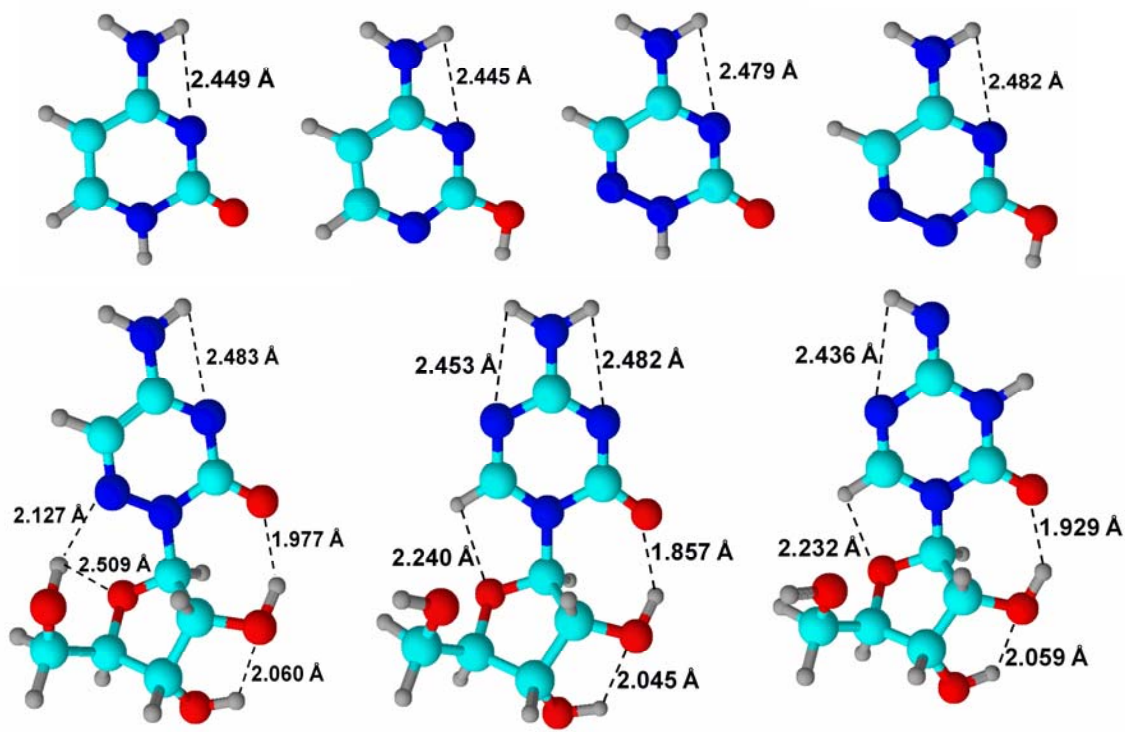

Figure 2. Structure of (a) cytosine, keto form, (b) cytosine, enol form, (c) 6-azacytosine, keto form, (d) 6-azacytosine, enol form, (e) 6-azacytidine, (f) 5-azacytidine and (g) 5-azacytidine, imino tautomer. Hydrogen bond distances are shown. Color code: grey, H; dark blue, N; red, O; light blue, C.

### 3.2. Core level XPS results.
### 3.2.1. Au 4f

The Au $4f_{7/2}$ core level spectra before and after adsorption of a monolayer of the present compounds from the liquid phase are shown in Figure 3. The spectrum of the clean surface consists of a doublet structure arising from the surface, peak *A* at energy 83.64 eV for Au(111) or 83.70 eV for Au(110), and bulk-derived peak *B* at 83.98 eV. After adsorption of all adsorbates from the liquid phase, the feature *A* of the surface state component decreased in intensity, and shifted by 50 meV to higher binding energy. This indicates that the adsorbed molecules are most likely chemisorbed rather than physisorbed on the gold surface. Chemisorption via the $N_{(3)}$ atom of the pyrimidine ring has been suggested for cytosine [40], and for related compounds such as cytidine and cytidine 5'-monophosphate[13]. It is also possible that the surface state intensity and energy vary due to a change of the reconstruction of the surface from the usual herringbone structure for Au(111), or the (1x2) reconstruction of Au(110). Since alteration of the reconstruction requires a chemical interaction, this interpretation leads to the same conclusion, that chemical bonding occurs. Kelly et al [19] found that the reconstruction was not lifted for cytosine evaporated onto Au(111), whereas Mansley et al [25] found that adsorption of cytosine from solution prevented the (1x2) reconstruction of Au(110).

On the (111) surface (see Fig. 3), all adsorbates induce a shift of the surface state of about 50 meV. However for the more open (110) surface, the vacuum deposited cytosine induces a larger shift of about 100 meV, while the solution deposited adlayer appears to shift the surface atom core levels to the value of the bulk Au atoms, i.e. about 300 meV with respect to the clean surface. The interaction with the (110) surface is therefore stronger in the sense that cytosine induces larger spectral changes.

The layer thickness has also been estimated from the inelastic mean free path of the photoelectrons, and the attenuation of the Au 4f signal, and the results are shown in Table 1. The effective thicknesses are similar to values obtained for other small bio molecules, such as peptides [10].

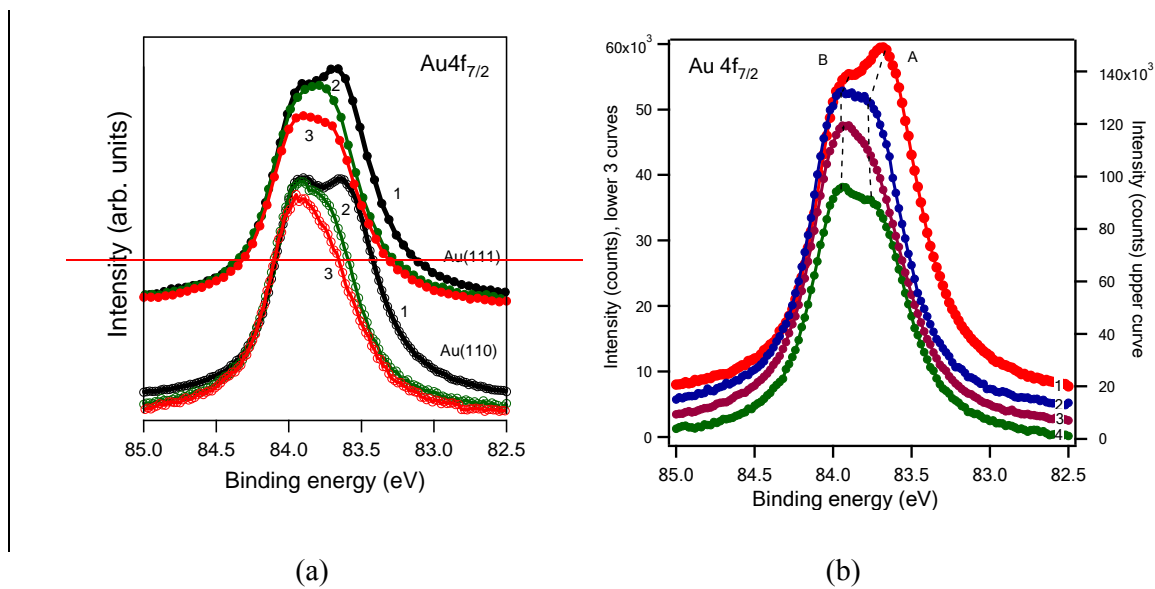

(a) (b)

Figure 3. Au $4f_{7/2}$ core-level spectra. (a) Comparison of adsorption of cytosine adsorbed on Au(111) (upper curves) and Au(110) (lower curves). (1): clean surface; (2): cytosine deposited from vacuum; (3): cytosine deposited from solution. (b) Comparison of adsorption on Au(111). (1) clean surface; (2) 6-azacytosine (3); 6-azacytidine; (4) 5-azacytidine. Photon energy: 120 eV. Intensity of the spectrum of the clean surface is shown on the right y axis; other spectral intensities are shown on the left axis. Curves have been offset for clarity.

| Sample/surface | adsorption mode | Au 4f attenuation | Effective thickness (Å) |
|---|---|---|---|
| cytosine, Au(111) | solution | 0.355 | 4.66 |
| cytosine, Au(111) | vacuum | 0.404 | 4.07 |
| cytosine, Au(110) | solution | 0.250 | 6.23 |
| cytosine, Au(110) | vacuum | 0.260 | 6.06 |
| 6-azacytosine, Au(111) | solution | 0.307 | 4.34 |
| 6-azacytidine, Au(111) | solution | 0.271 | 4.36 |
| 5-azacytidine, Au(111) | solution | 0.347 | 3.54 |

Table 1. Au 4f intensity attenuation (intensity for the adsorbate covered surface divided by the intensity measured for the clean surface) due to adsorption, and effective thickness.

*3.2.2. Core level spectra of cytosine*

The experimental C 1s, N 1s, and O 1s photoemission spectra of monolayer coverages of cytosine adsorbed from solution and from vacuum on Au(110) and Au(111) are presented in Figure 4, compared with the gas phase spectra [8]. At the temperature used here, three tautomers are present in the gas phase, and on the surface it is expected that they will rapidly achieve a new equilibrium population ratio. The individual gas phase tautomers have similar band widths for the C and N 1s spectra, and only the O 1s spectrum shows a single peak whose energy varies depending on which tautomer is present. The C 1s spectra of cytosine deposited on both surfaces by both methods consist of a band of peaks from 284.6 to 287.7 eV, width 3.1 eV, compared with the gas phase spectra which stretch from 290.6 to 293.9 eV, width 3.3 eV (excepting only the $C_{(2)}$ 1s state of the imino-keto tautomer, which lies outside this range. Refer to Supplementary materials Fig S2 for the nomenclature). The bandwidths are almost the same, and the average shift from the gas phase is 6.1 eV, due largely to the work function (i.e. different reference levels: the vacuum level in the gas phase, the Fermi level for the solid state), but also including effects due to screening of the hole states on the surface. Similarly the N 1s spectra have peaks between 398.2 and 399.75 eV on the surface and 404.5 and 406.1 eV in the gas phase, with similar bandwidths, and a shift of 6.3 eV. Thus we expect the O 1s peaks will be shifted by about this energy. The present experimental O 1s binding energies were 530.8 eV in 3 cases, and 531.3 eV for vacuum deposition of cytosine on Au(110). For a shift of 6.4 eV between the solid and gas phases, the energies in the gas phase are about 537.2 and 537.8 eV. This excludes the presence of an enol tautomer on the surface, as its O 1s gas phase binding energy is 539.4 eV. Tautomers 1 (amino-keto) and 3 (imino-keto) of cytosine have binding energies of 536.5 and 537.3 eV, so these are candidates for the tautomeric forms which are present on the surface.

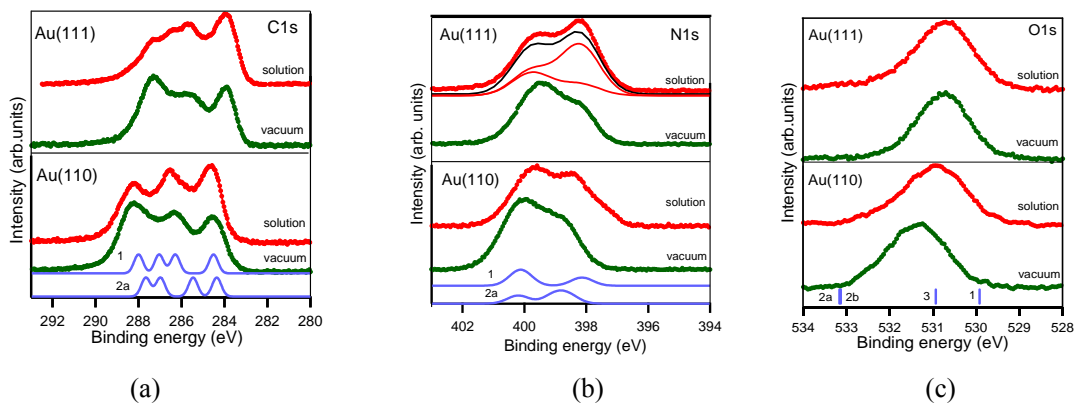

(a)  (b)  (c)

Figure 4. Cytosine adsorbed on Au(111) and Au(110) from solution and by evaporation, compared with the gas phase spectrum (1, 2a, 2b, 3 are theoreticaly calculated cytosine tautomers [8]). (a) C 1s spectrum. (b) N 1s spectrum. (c) O 1s spectrum.

|             | Cytosine on Au(110) surface | | Cytosine on Au(111) surface | |
|---|---|---|---|---|
| Atomic site | Vacuum dep. | Solution dep. | Vacuum dep. | Solution dep. |
| C 1s | 288.5 | 287.9 | 287.9 | 287.8 |
|      | 286.5 | 286.2 | 285.9 | 286.0 |
|      | 284.7 | 284.3 | 284.2 | 284.2 |
| N 1s | 400.0 | 399.7 | 399.5 | 399.8 |
|      | 398.8 | 398.4 | 398.3 | 398.2 |
| O 1s | 531.4 | 530.9 | 530.7 | 530.7 |

Table 2. Binding energy (eV) of cytosine deposited from vacuum and solution on Au(110) and Au(111) surfaces.

The hypothesis of an imino-keto tautomer after vacuum deposition on Au(110) can be tested by checking the carbon spectrum. In the gas phase, this tautomer has a $C_{(2)}$ 1s peak at 1.2 eV higher binding energy than all other C 1s states of all tautomers [8]. This peak is in fact absent from the spectrum, so we exclude the formation of this tautomer. To explain the differences in the spectra, other hypotheses are necessary. We can exclude deprotonation as a possibility, because the thermal desorption spectra [31,32] show that cytosine desorbs intact. We conclude the chemical shifts are due to intermolecular hydrogen bonding and/or interaction with the surface.

It is known from gas phase studies that the N 1s core level can shift by about 0.5 eV in amino acids, depending on whether there is an internal OH…N hydrogen bond, or not [41,42]. Similar or even larger shifts have been predicted for the O 1s core levels involved in hydrogen bonding [43], but they have not been positively identified in free molecules because of the large intrinsic width of the O 1s core levels. While these shifts are substantial, they cannot alone explain the changes in the N 1s spectra. Two peaks are present in all N 1s spectra with a separation of about 1.5 eV and they change their relative intensity, depending on which gold surface and which deposition technique is used. In the gas phase, the two peaks are assigned to amino nitrogen at higher binding energy, and imino nitrogen at lower binding energy. However we have established that imino tautomers are not present, because the O 1s spectrum does not show the large changes expected.

We therefore conclude that the N 1s shifts are due to interaction of the amino nitrogen groups with the surface. Previous calculations for the dipeptide glycyl-glycine show that an amino group can interact with the surface via the hydrogen atom of the amino group, leading to a N 1s shift of 1.25 to 1.5 eV [44] (for the species labeled AN2 and AN4 in that work.) On Au(111) and Au(110), the molecule is present mainly as the amino-keto tautomer (Fig. 1a), in a hydrogen bonded network [21], which we label species I. The N 1s peaks reflect this chemical state, with the higher energy (amino) peak more intense than the lower energy (imino) peak. However the ratio is not the expected stoichiometric ratio of 2:1, and even taking account of shifts due to hydrogen bonding, the spectrum cannot be fitted with peaks having this ratio. Thus we conclude that a second form with bonding to the surface via amino hydrogen is present, which we label species II. For vacuum deposition on Au(111) and Au(110) this species is the minority form, while for solution deposition on Au(111), it is the majority species. In fig. 4, we have fitted the spectra with a sum of two spectra corresponding to I and II: for I the intensity ratio is 2:1 representing the amino: imino stoichiometry; and for II the intensity ratio is 1:2, representing the amino:(imino+shifted amino) intensity ratio. We estimate the ratio of the two species to be I:II = 0.31:0.69.

The C 1s spectrum shows some differences with respect to the gas phase, although the separation in binding energy of $C_{(2)}$ and $C_{(5)}$ remains nearly the same. The shifts of the other two core levels, $C_{(4)}$ and $C_{(6)}$, are attributed to indirect effects of hydrogen bonding and chemical interaction with the surface. The N 1s and C 1s spectra of vacuum deposited cytosine are very similar for Au(111) and (110), although the O 1s spectrum shows a significant difference in energy between the two surfaces. This may be due to the effect of changes in the (1x2) reconstruction of the Au(110) surface, but we do not have a detailed explanation. For solution deposition, the N 1s spectra indicate a much higher proportion of species II on Au(111), and an increased proportion on Au(110). Corresponding spectral changes occur in the C 1s spectra, which (a little surprisingly) resemble the gas phase spectrum more closely than the evaporated samples.

*3.2.3. Core level spectra of cytosine derivatives*

The experimental O 1s, N 1s, and C 1s photoemission spectra of monolayer coverages of 6-azacytosine, 6-azacytidine and 5-azacytidine on Au(111) are presented in Figure 5, together with the calculated free molecular spectra. The theoretical energies are summarized in Table 3.

The O 1s spectrum of 6-azacytosine shows a shoulder A at 532.5 eV on the main peak B at 531.0 eV, assigned to $O_{(2)}$ of the azacytosine ring. We interpret the spectrum as evidence of the presence of tautomers [8,45] on the surface, in particular the fitted peak B is assigned to the keto form and peak A to the imino form of 6-azacytosine, by analogy with cytosine in the gas phase. The ratio of peak B:A intensities, and therefore populations, is about 2.8 : 1. Peak A is not well resolved, and the experimental difference in binding energy, 1.6 eV, is in qualitative agreement with a theoretical value of 2.3 eV, see curves a and a' in fig. 5(i).

The N 1s spectrum of 6-azacytosine was fitted with three peaks, A, B and C. The theoretical curves in Fig. 4(ii) show that peaks A and C are associated with the keto tautomer, while peak B is assigned to the enol tautomer. The estimated population ratio (A+C) : B is 2.8 : 1 ±0.2, in good agreement with the value estimated from the O 1s spectrum. We note that the explanation of the N 1s peak intensity ratio given for cytosine (molecule-surface interaction) would be valid, if the only data available were the nitrogen core level spectrum. However in this case, both the O and N 1s spectra point to tautomerism. The C 1s spectra do not provide much further information: the peaks A/A' and B are qualitatively consistent with the theoretical spectra, although the intensity ratios are not quantitative, and peak C at 284.7 eV is assigned to adventitious carbon.

The O 1s spectrum of 6-azacytidine is dominated by peak A, due to the hydroxyl C-OH and ether C-O-C oxygen atoms of the ribose ring, while the peak B is assigned to the keto oxygen of the azacytosine ring. Since the O 1s peak of the enol tautomer overlaps the peak A, it is not possible to determine whether this tautomer is present. However, the N 1s spectrum is very similar to that of 6-azacytosine, so we believe that both compounds have the same tautomer populations of the azacytosine moiety. The C 1s spectrum differs due to the carbon atoms of the ribose ring: there is additional intensity in the central peak B of 6-azacytidine due to these carbon atoms, all of which are bonded to oxygen, with corresponding chemical shifts.

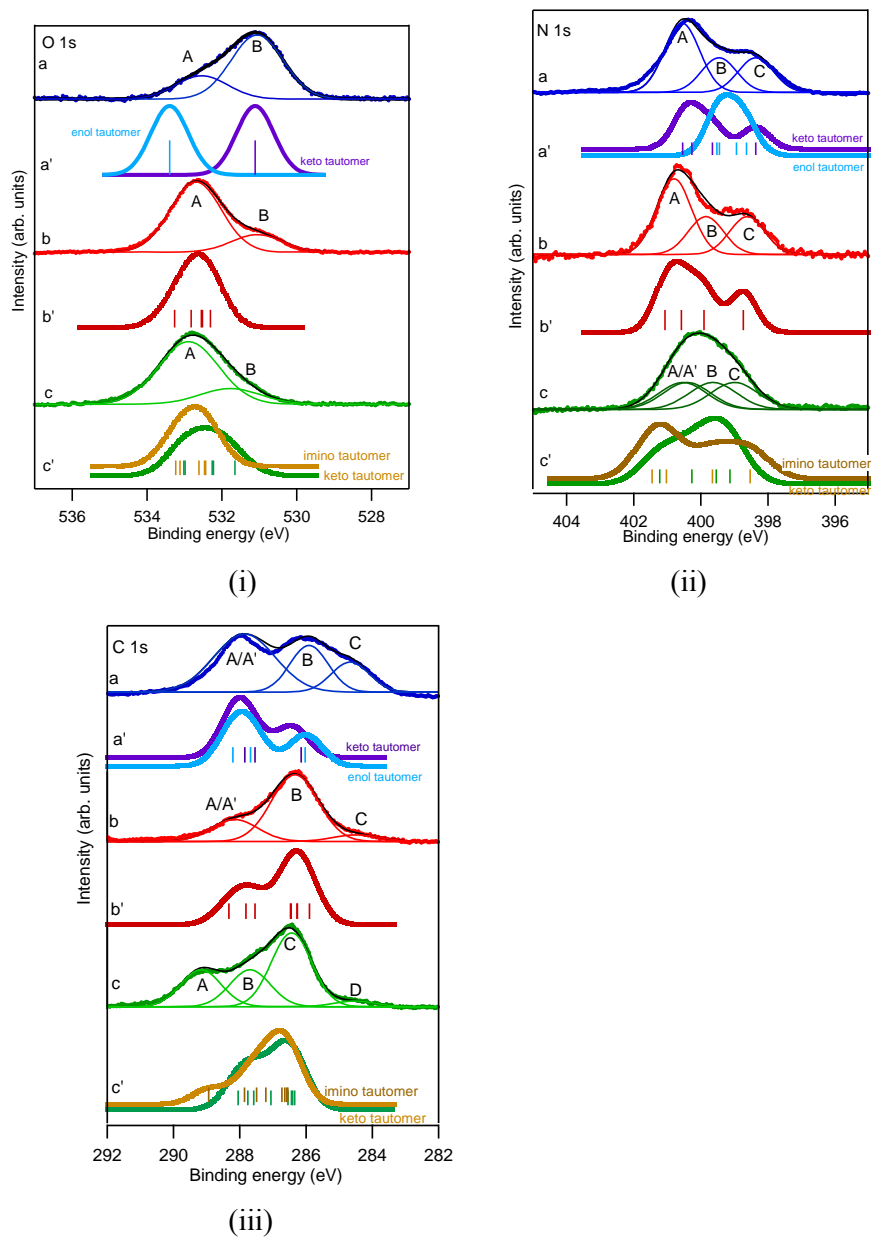

Figure 5. (i) O, (ii) N and (iii) C 1s photoemission spectra of (a) 6-azacytosine, (b) 6-azacytidine, and (c) 5-azacytidine. Experimental spectra (a-c) are of molecules adsorbed on Au(111) and calculated spectra (a'-c') are of free molecules. Calculated spectra (referred to the vacuum level) are shifted to lower energy to align with the experimental data (referred to the Fermi level) by: O 1s: (a') –2.8 eV, (b') –2.18 eV, (c') –2.55 eV; N 1s: (a') –5 eV, (b') –5 eV, (c') –4 eV; C 1s: (a') –4.9 eV, (b') – 5.2 eV, (c') –5.2 eV.

The O 1s spectrum of 5-azacytidine shows a single broad peak, with a slight shoulder B at lower binding energy, shifted to higher energy than the corresponding peaks in 6-azaC and 6-azacytidine. The calculations show that this is expected for an imino tautomer, so we hypothesise that this compound exists predominantly as a keto-imino tautomer, similar to tautomer 3 of cytosine [8,45]. It is also possible that it is an enol tautomer, as the enol peak is expected at the energy of the main peak A, due to hydroxyl and ether oxygen atoms. However the C 1s spectrum, Fig. 5(iii), allows discrimination of the tautomer type. The carbon spectrum is unusually broad, with a peak A at higher binding energy than for the other compounds. This is correctly modeled by the calculations for the imino tautomer, and is similar to the situation for cytosine in the gas phase [8,45]: the carbon atom $C_{(2)}$ of the imino tautomer shows an unusually high binding energy, as was predicted by theory and verified by experiment.

The agreement with the theoretical N 1s spectrum of 5-azacytidine is not good, and the calculations predict a broader band than that found in the experiment. We speculate that this is due to interaction with the surface of the amino and imino hydrogen atoms, leading to shifts to lower binding energy, analogous to the effect described above for cytosine.

**Table 3.** Comparison between the simulated and measured vertical core ionization potentials (eV) of 6-azaC, 6- azacytidine, 5- azacytidine.

| Atomic site | 6-azacytosine | | | 6-azacytidine | | 5-azacytidine | | |
|---|---|---|---|---|---|---|---|---|
| | LB94/et-pVQZ | | Exp. | LB94/et-pVQZ | Exp. | LB94/et-pVQZ | | Exp. |
| | Keto-form | Enol-form | | Global (anti) | | Global (anti) | Imino tautomer | |
| O-K | Shifted -2.8 eV | | | Shifted -2.18 eV | | Shifted -2.55 eV | | |
| $O_{(2)}$ | 531.11 | 533.4 | 531.06 532.54 | 532.31 | 531.06 | 531.65 | 532.46 | 531.77 |
| $O_{(2')}$ | | | | 532.82 | | 532.26 | 532.61 | |
| $O_{(3')}$ | | | | 532.55 | 532.7 | 532.23 | 532.43 | 532.89 |
| $O_{(4')}$ | | | | 533.27 | | 533.02 | 533.23 | |
| $O_{(5')}$ | | | | 532.52 | | 532.98 | 533.12 | |
| | | | | | | | | |
| N-K | Shifted -5 eV | | | Shifted -5 eV | | Shifted -4 eV | | |
| $N_{(1)}$ | 400.54 | 398.94 | 400.55 | 401.07 | 400.8 | 401.23 | 401.45 | 399 |
| $N_{(3)}$ | 398.36 | 398.64 | 398.35 | 398.74 | 398.62 | 399.14 | 401.02 | 399.64 |
| $N_{(4)}$ | 399.65 | 399.52 | 399.45 | 399.9 | 399.85 | 400.27 | 398.53 | 400.54 |
| $N_{(5)}$ | | | | | | 399.54 | 399.65 | 400.41 |
| $N_{(6)}$ | 400.27 | 399.45 | 400.55 | 400.58 | 400.8 | | | |
| | | | | | | | | |
| C-K | Shifted -4.9 eV | | | Shifted -5.2 eV | | Shifted -5.2 eV | | |
| $C_{(2)}$ | 288.15 | 288.21 | 287.85 | 288.33 | 288.16 | 288.05 | 288.94 | 289.17 |
| $C_{(4)}$ | 287.84 | 287.68 | | 287.81 | | 287.76 | 287.49 | |
| $C_{(5)}$ | 286.45 | 286.03 | 285.9 | 286.45 | 286.34 | | | |
| $C_{(6)}$ | | | | | | 287.06 | 287.21 | 287.69 |
| $C_{(1')}$ | | | | 287.54 | 288.16 | 287.58 | 287.86 | |
| $C_{(2')}$ | | | | 286.47 | | 286.41 | 286.64 | |
| $C_{(3')}$ | | | | 286.26 | 286.34 | 286.34 | 286.55 | 286.44 |
| $C_{(4')}$ | | | | 286.27 | | 286.55 | 286.73 | |
| $C_{(5')}$ | | | | 285.9 | | 286.44 | 286.58 | |

In the C 1s spectra of all molecules, the weak feature *D* is due to carbon containing impurities, and the BE of this peak suggests that they are hydrocarbons. The C 1s spectrum of 6-azacytosine shows some evidence of this carbon contamination, with a peak at 284-285 eV (about 20% of maximum intensity). All three carbon atoms in this molecule are bonded to one, two or three electronegative partners, and so have high binding energies. In 6-azacytidine, the peak A at about 288.3 eV corresponds to the same peak in 6-azacytosine, assigned to the urea carbon $C_{(2)}$, and with $C_{(4)}$ and $C_{(5)}$ at 0.5 and 1.9 eV lower binding energy respectively. 6-azacytidine contains 8 carbon atoms in total, and these mainly contribute to the strongest peak C. The overall agreement between theory and experiment is satisfactory.

*3.2. NEXAFS spectra*

The molecular adsorption geometry of the monolayer films was analyzed using NEXAFS spectroscopy. Figure 6 shows the N and O K-edge NEXAFS spectra of cytosine, measured at normal and grazing incidence of the photon beam with respect to the surface. The sharp, low energy peaks around 400 eV for nitrogen and 532 eV for oxygen are due to transitions to π type antibonding orbitals, and the broad maxima at 408-415 and about 540 eV are due mainly to transitions to σ transitions [46]. The low energy region is shown in more detail in Figure 7 for the grazing incidence spectra on Au(110), compared with theory; the spectra from Au(111) are similar.

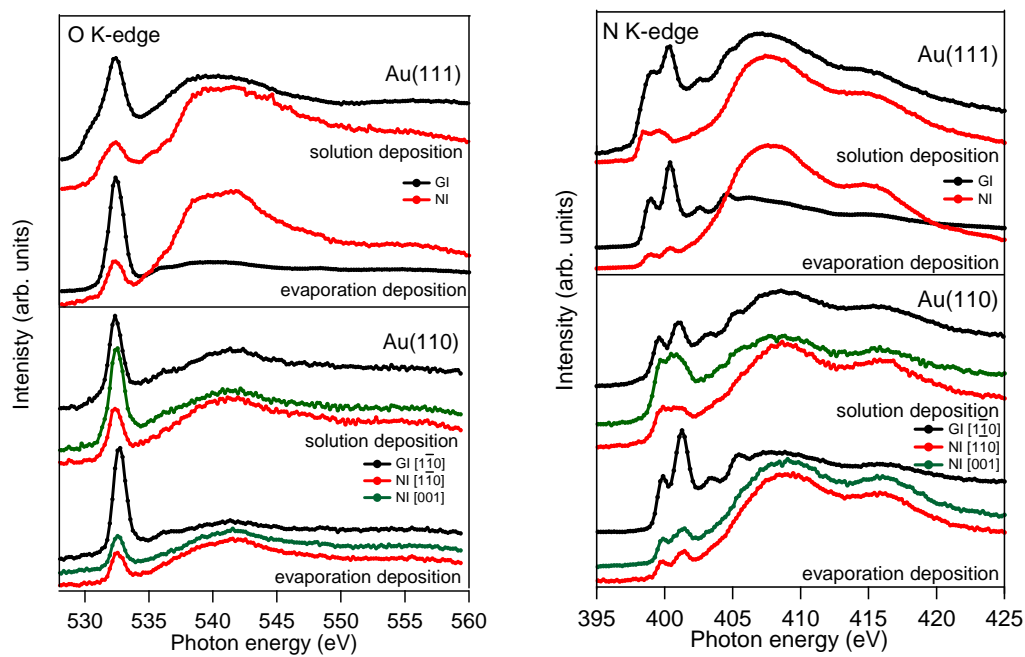

Figure 6. O and N K-edge NEXAFS spectra of cytosine on Au(111) and Au(110). The method of preparation, the substrate and incidence (GI: Grazing Incidence; NI: Normal Incidence) are indicated. For Au(110), the azimuth of the plane of incidence is also indicated.

All spectra of cytosine show much stronger anisotropy between grazing and normal incidence for the vacuum deposited samples, indicating that the molecular plane is oriented nearly parallel to the surface. The energies of the spectral features do not depend on the substrate orientation, and the same N and O edge spectra are observed on both surfaces, with different intensity ratios for peaks of different symmetry. From the photoemission data above, we know that two species labelled I and II are present on the surface. The residual intensity of the π type transitions at normal incidence indicates that one or both of these species is not quite parallel to the surface. There may also be dynamic fluctuations, either in geometry or between the two phases, as the surface structure is fluid.

The energies of the peaks close to threshold can be interpreted in the light of the gas phase spectrum [45]. In Figure 7, three peaks are clearly visible and correspond to similar peaks in the gas phase. The peaks B and C have the same energy on the surface and so are unperturbed. Peak A, due to transitions from $N_{(3)}$ 1s to antibonding states $V_1$, is however shifted to higher energy, indicating chemical interaction with the surface. The shift in photon energy of this transition is qualitatively consistent with a model for cytosine in an electrochemical cell [22] which predicts that cytosine interacts with Au via the $N_{(3)}$ nitrogen; resonances due to other nitrogen atoms are less perturbed. Resonance B is due to transitions from $N_{(7)}$ and $N_{(1)}$ 1s to antibonding states $V_1$ and 3s, while feature C has been tentatively assigned to transitions of type $N_{(7)}$ 1s → 3p, $N_{(3)}$ 1s → $V_2$ and $N_{(1)}$ 1s → $V_2$. The presence of multiple contributions to this feature explains why it does not shift significantly, in contrast to peak A.

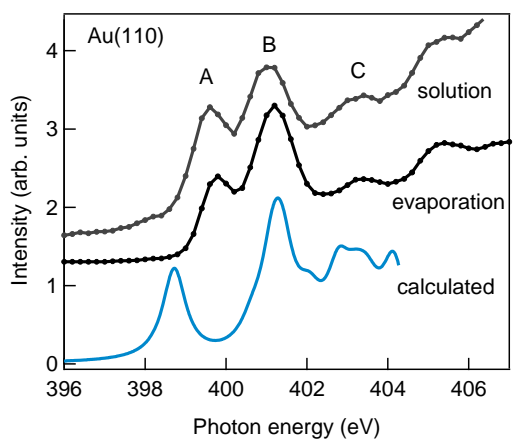

Figure 7. Low energy region of the N K edge NEXAFS spectra of cytosine at GI geometry. Upper curve: sample prepared from solution; center curve: sample prepared by evaporation; lower curve: theoretical spectrum of tautomer 1 in the gas phase [45]. The theoretical energies have been convoluted with Lorentzian functions and weighted by their pole strengths. Only states below threshold are included in the theoretical spectrum.

For the solution deposited samples, in which phase II dominates, the GI/NI anisotropy is much reduced. Indeed on the (110) surface, both the oxygen and nitrogen K edge spectra have the highest π/σ intensity ratio for incidence along the [001] direction, indicating that the molecular plane is oriented along this direction. The O 1s spectra show similar shapes, except for the solution deposition, where normal and grazing incidence spectra have an extra shoulder at 1 eV below the main π resonance. Since the XPS spectra above are consistent with the presence of two species, a quantitative analysis is not possible.

The N K edge NEXAFS spectral shapes show some differences between the vacuum and solution deposited samples. Both sets of data show two main peaks, followed by a weaker structure at about 402.5 eV, and broad σ resonances at higher energy. However the spectra at normal incidence on both surfaces show substantial differences, indicating that the chemical state of the dominant species is different for the two preparations. This is what we expect for a change of chemical state centered on one of the three nitrogen atoms, in particular $N_{(3)}$.

In Figure 8, the O and N K-edge NEXAFS spectra of 6-azacytosine, 6-azacytidine and 5-azacytidine are shown. 6-azacytosine shows a weak angular effect at both K edges, and the π resonances are stronger at grazing incidence. Since the dipole moments of the π orbitals are oriented perpendicular to the molecular plane, this implies that the molecular plane is preferentially oriented parallel to the surface. The resonances are still relatively strong at normal incidence, so the molecules are not strongly oriented, and may be either at a fixed angle, or disordered. The O 1s photoemission spectra above for cytosine indicate that more than one tautomer is present, and they may have different orientations.

The N K edge NEXAFS spectrum of 6-azacytidine is the same within experimental error as that of 6-azacytosine, indicating that the addition of the ribose moiety does not alter the preferred molecular orientation on the surface. The O K edge NEXAFS spectrum is substantially different because of the additional oxygen atoms, all with saturated bonds, which add intensity to the σ transition region.

The ratio of intensities at the N K edge shows much stronger contrast for 5-azacytidine than for 6-azacytosine and 6-azacytidine, in particular the π resonances are nearly extinguished at normal incidence. Thus the simple displacement of the imino nitrogen from one position in the azacytosine ring to another produces a much more parallel geometry of the adsorbate.

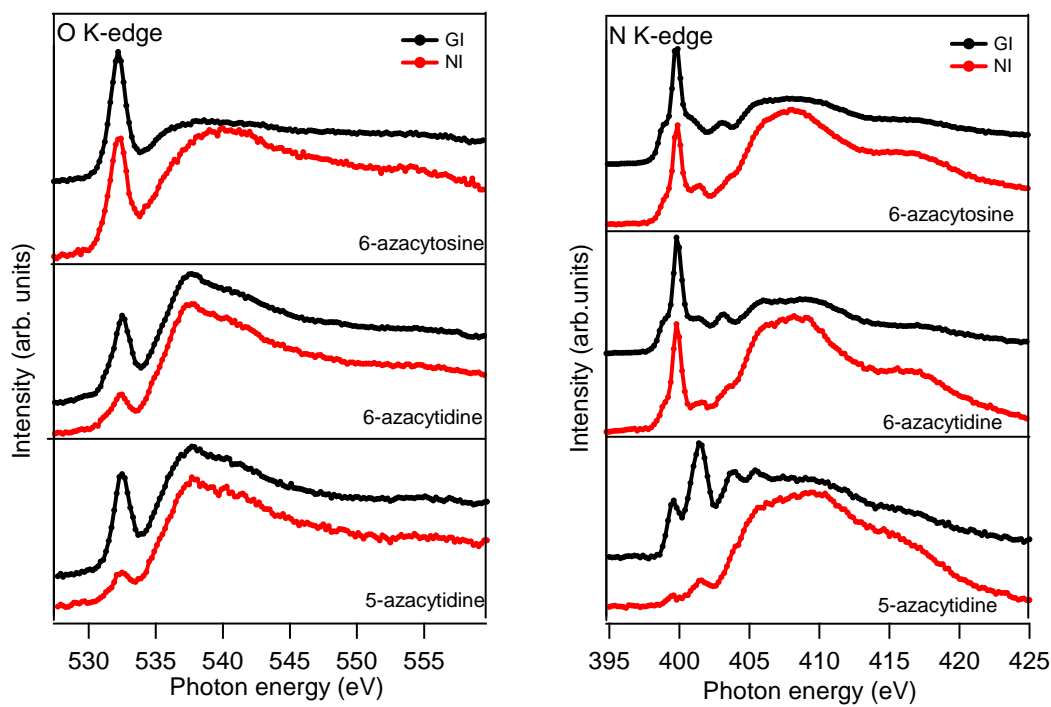

Figure 8. O, N K-edge NEXAFS spectra of 6-azacytosine, 6-azacytidine, 5-azacytidine.

## 3.4. Adsorption model

The results described above can be combined to construct a model for the adsorption of these molecules on gold. The Au 4f spectra showed chemical shifts of the surface peak for all molecules, which is an indication of a chemical interaction, rather than interaction via weak dispersion forces. In the gas phase, cytosine exists as three tautomers, but on the surface, a single keto tautomer is present. Adsorbed cytosine has been extensively studied, and for example STM studies of samples evaporated in vacuum reported a single species on the surface [19,21], whereas our XPS data clearly show the existence of two states. The discrepancy may be due to the lack of chemical sensitivity of STM, where molecules in different chemical states may give similar images. It could also be due to the different temperatures of the experiments - STM at low temperature, and the present work at room temperature. The two chemical states may have different free energies as a function of temperature, and the higher energy state may be missing at low temperature. We consider this explanation less likely as the two chemical states appear to be present when cytosine is adsorbed from solution, but with a different population ratio. Thus the formation of the states does not seem to depend primarily on temperature but on preparation method.

We tentatively identify our chemical state I with the flat-lying, hydrogen bonded species described in the STM work. Although the adsorption energy, as measured by thermal desorption or predicted theoretically, may be low, this species interacts strongly with the surface in the sense that the electronic structure is significantly perturbed. This is particularly evident in the N spectra, where the N 1s spectra show some energetic shifts with respect to the gas phase, and the N K edge NEXAFS spectra show a clear shift of a peak assigned to $N_{(3)}$ in the gas phase. We conclude that this species interacts with the surface via $N_{(3)}$, as well as being hydrogen bonded to its neighbors. This result is consistent with the calculations of Bogdan and Morari [14], but inconsistent with those of other authors [18,19,21].

State II is the majority species in solution prepared samples, and is tilted at a large angle to the surface and has the molecular plane oriented approximately along the [100] direction. We identify it with the phase reported in STM studies in electrochemical cells [22,28], and by optical studies in solution [26]. In the last study, the measurements were performed in situ, and the molecules were found to be perpendicular to the surface and oriented along [1 $\bar{1}$ 0], whereas we found an intermediate angle. However the present measurements were performed ex situ, and some molecular rearrangement may occur on transfer to vacuum.

This state is a keto tautomer and is also bound via $N_{(3)}$ to the surface. The N 1s XPS results also provide evidence of interaction of the amino nitrogen $N_{(7)}$ with the surface, as the core level energy is strongly shifted to lower binding energy. For steric reasons, a cytosine molecule bonded edge-on to the surface via $N_{(3)}$ and $N_{(7)}$ can be expected to have the oxygen atom close to the surface as well, but we did not find spectroscopic evidence of chemical interaction. This model is similar to one proposed earlier [40].

On changing the structure of cytosine to 6-azacytosine, the tautomeric equilibrium is changed, and there is no longer a single tautomer present, but two species, identified as keto and enol tautomers. Both the O and N 1s spectra support this model. 6-azacytidine adsorbs via the azacytosine ring, and the ribose ring does not strongly influence the chemisorption. This is similar to the situation for the closely related molecule deoxycytidine on Au(111) [14].

For 5-azacytidine, the oxygen and carbon spectra provided evidence that the tautomeric equilibrium was shifted towards the imino form on the surface. We interpret the N 1s spectra as indicating that the imino and amino nitrogen atoms were interacting with the surface via their hydrogen atoms.

While 5-azacytidine is very nearly parallel to the surface, 6-azacytidine and 6-azacytosine are only partly oriented. There are at least two possible mechanisms for stronger orientational effects, either increased interaction between the surface and the π orbitals of the azacytosine ring, or increased intermolecular interaction, which promotes a flat geometry. For the imino tautomer there are fewer possibilities for forming multiple hydrogen bonds, so it appears that electronic effects and interaction with the surface dominate.

With regard to the ribose moiety, this does not appear to influence the adsorption geometry strongly. This is in keeping with the expectation that this structural unit does not interact strongly with the metal surface, and that bonding proceeds via the azacytosine groups.

## 4. Conclusions

In this work we have examined the interaction with gold surfaces of cytosine and three relatively complex derivatives, of pharmacological interest. Cytosine adsorbs as a keto tautomer, but in two chemical states depending on the mode of deposition. Adsorption from the vapor favors the flat lying state I, which in any case also interacts with the surface via $N_{(3)}$. Deposition from solution favors a second state II, which is tilted at an angle to the surface and displays a different N 1s core level spectrum. This state interacts with the surface via $N_{(3)}$ and $N_{(7)}$, and is identified with similar phases observed in electrochemical cells and in solution.

The derivatives are present on the surface as mixtures of tautomers. Addition of a ribose ring to make azacytidine does not influence the chemisorption mode, which in any case occurs via the heterocycle ring.

**Acknowledgements**

We gratefully acknowledge the assistance of our colleagues at Elettra for providing good quality synchrotron light. The Materials Science Beamline is supported by the Ministry of Education of the Czech Republic under Grant No. LC06058. We would like to kindly thank I. Alexeeva and L. Palchykovska from the Institute of Molecular Biology and Genetics, National Academy of Science of Ukraine for providing the samples of 6-azacytidine and its derivatives. The Abdus Salam International Center for Theoretical Physics is acknowledged for financial support through a STEP fellowship. FW and MA acknowledge the National Computational Infrastructure (NCI) at the Australian National University under the Merit Allocation Scheme (MAS) and MA acknowledges the Swinburne University Postgraduate Research Award (SUPRA).

Supplementary data.

Table S1. Comparison of selected geometric parameters of 5-azaC and 6-azaC with available results.

| Parameters | 5-azaC | 6-azaC | |
|---|---|---|---|
| | This work | This work | Singh et al.[a] |
| $R_5$ (Å) | 7.331 | 7.470 | 7.456 |
| $R_6$ (Å) | 8.129 | 8.163 | 8.168 |
| $\angle C_{(4)}-N_{(5)}-C_{(6)}$ (°) | 114.36 | 120.25[b] | 120.80 |
| $\angle N_{(5)}-C_{(6)}-N_{(1)}$ (°) | 124.01 | 117.87[c] | 117.60 |
| $\chi = \angle O_{(4')}-C_{(1')}-N_{(1)}-C_{(2)}$ (°) | -173.52 | -162.65 | |
| $\gamma = \angle C_{(3')}-C_{(4')}-C_{(5')}-O_{(5')}$ (°) | 54.27 | 61.07 | |
| $P$ (°) | 180.38 | 184.01 | |
| $v_{max}$ | 30.52 | 29.43 | |
| $<R^2>$ (a.u.) | 4077.524 | 4041.658 | |
| $\mu$ (D) | 6.11 | 6.94 | |
| Total Energy ($E_h$) | -907.487546 | -907.443716 | |
| ZPE (kcal.mol$^{-1}$) | 142.029 | 141.682 | |
| T.E +ZPE($E_h$) | -907.261208 | -907.217932 | |
| Sugar type | C3'-exo | C3'-exo | |

[a] Crystal structure, P. Singh, D. J. Hodgson, *Biochemistry,* **1974,** *13* (26), 5445-5452.
[b] $\angle C_{(4)}-C_{(5)}-N_{(6)}$ (°).
[c] $\angle C_{(5)}-N_{(6)}-N_{(1)}$ (°).

**Table S2.** Intramolecular C(O)–H⋯O(N) distances of the nucleoside derivatives [a].

| (C(O)–H⋯O(N)) bond (Å) | 5-azaC | 6-azaC |
|---|---|---|
| **Sugar -Base(SB)** | | |
| $C_{(6)}$– H⋯ $O_{(4')}$ | 2.240 | |
| $C_{(6)}$– H⋯ $O_{(5')}$ | 3.038 | |
| $C_{(1')}$– H⋯ $O_{(2)}$ | 2.604 | 2.457 |
| $C_{(2')}$– H⋯ $O_{(2)}$ | 3.162 | 3.345 |
| $C_{(1')}$– H⋯ $N_{(1)}$ | 2.081 | 2.067 |
| $C_{(2')}$– H⋯ $N_{(1)}$ | 2.593 | 2.576 |
| $O_{(2')}$– H⋯ $O_{(2)}$ | 1.857 | 1.977 |
| $O_{(5')}$– H⋯ $N_{(6)}$ | | 2.127 |
| **Sugar-Sugar (SS)** | | |
| $C_{(1')}$– H⋯ $O_{(2')}$ | 2.560 | 2.544 |
| $C_{(1')}$– H⋯ $O_{(3')}$ | 3.104 | 3.160 |
| $C_{(1')}$– H⋯ $O_{(4')}$ | 2.064 | 2.059 |
| $C_{(2')}$– H⋯ $O_{(2')}$ | 2.082 | 2.079 |
| $C_{(2')}$– H⋯ $O_{(3')}$ | 3.341 | 3.336 |
| $C_{(2')}$– H⋯ $O_{(4')}$ | 2.880 | 2.902 |
| $C_{(2')}$– H⋯ $O_{(5')}$ | 2.586 | 2.739 |
| $C_{(3')}$– H⋯ $O_{(2')}$ | 2.909 | 2.939 |
| $C_{(3')}$– H⋯ $O_{(3')}$ | 2.075 | 2.076 |
| $C_{(3')}$– H⋯ $O_{(4')}$ | 3.287 | 3.290 |
| $C_{(3')}$– H⋯ $O_{(5')}$ | 2.840 | 3.052 |
| $C_{(4')}$– H⋯ $O_{(3')}$ | 2.405 | 2.401 |
| $C_{(4')}$– H⋯ $O_{(4')}$ | 2.059 | 2.063 |
| $C_{(4')}$– H⋯ $O_{(5')}$ | 3.336 | 3.380 |
| $C_{(5')}$– $H_a$⋯ $O_{(4')}$ | 2.634 | 2.689 |
| $C_{(5')}$– $H_a$⋯ $O_{(5')}$ | 2.088 | 2.081 |
| $C_{(5')}$– $H_b$⋯ $O_{(4')}$ | 3.368 | 3.369 |
| $C_{(5')}$– $H_b$⋯ $O_{(5')}$ | 2.092 | 2.032 |
| $O_{(3')}$– H⋯ $O_{(2')}$ | 2.045 | 2.060 |
| $O_{(5')}$– H⋯ $O_{(4')}$ | 3.748 | 2.509 |

[a] Based on the B3LYP/6-311++G** level of theory and applying the 2.8 Å cut off criterion.

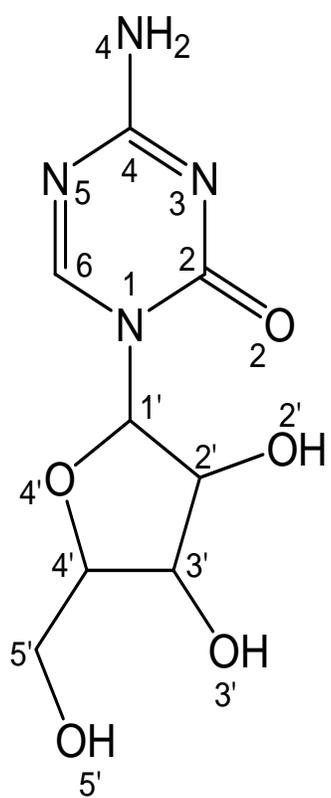 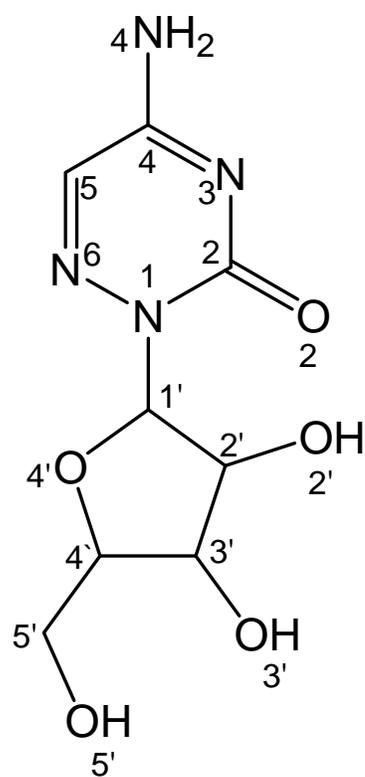

**5-Azacytidine**  **6-Azacytidine**

Fig. S2 Nomenclature of 5-azacytidine and 6-azacytidine.

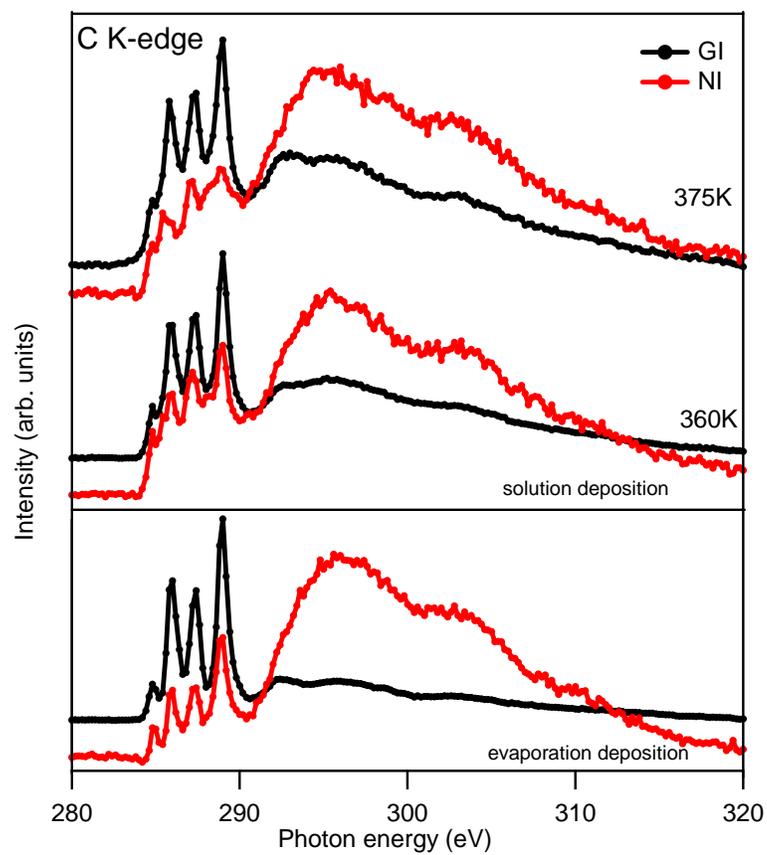

Fig.S2 C K-edge NEXAFS spectra of cytosine on Au(111)